\documentclass[twocolumn,showpacs,preprintnumbers,amsmath,amssymb]{revtex4}

\usepackage{graphicx}
\usepackage{dcolumn}
\usepackage{bm}

\begin{document}


\title{Bias current dependence of spin accumulation signals in a silicon channel detected by a Schottky tunnel contact}

\author{Y. Ando,$^{1,2}$ K. Kasahara,$^{1}$ K. Yamane,$^{1}$ Y. Baba,$^{1}$ Y. Maeda,$^{1}$ Y. Hoshi,$^{3}$ K. Sawano,$^{3}$ M. Miyao,$^{1}$ and K. Hamaya$^{1,4}$\footnote{E-mail: hamaya@ed.kyushu-u.ac.jp}}

\affiliation{$^{1}$Department of Electronics, Kyushu University, 744 Motooka, Fukuoka 819-0395, Japan}%
\affiliation{$^{2}$INAMORI Frontier Research Center, Kyushu University, 744 Motooka, Fukuoka 819-0395, Japan.}%
\affiliation{$^{3}$Advanced Research Laboratories, Tokyo City University, 8-15-1 Todoroki, Tokyo 158-0082, Japan}
\affiliation{$^{4}$PRESTO, Japan Science and Technology Agency, Sanbancho, Tokyo 102-0075, Japan}%

%

\date{\today}
\begin{abstract}
We study the electrical detection of spin accumulation at a ferromagnet-silicon interface, which can be verified by measuring a Hanle effect in three-terminal lateral devices. The device structures used consist of a semiconducting Si channel and a Schottky tunnel contact. In a low current-bias region, the Hanle-effect curves are observed only under forward bias conditions. This can be considered that the electrical detectability at the forward-biased contact is higher than that at the reverse-biased contact. This is possible evidence for the detection of spin-polarized electrons created in a Si channel. 
\end{abstract}
\maketitle
Electrical injection and detection of spin polarized electrons in semiconductors have been verified by measuring a Hanle effect using two- or three-terminal lateral devices recently.\cite{Lou,Appelbaum,Tran,Dash,Sasaki2} In GaAs-based devices, two different features have been reported as follows. One was seen in epitaxial Fe/GaAs devices with Schottky tunnel contacts, reported by Lou {\it et al}.\cite{Lou} They showed that evident Hanle-effect curves are observed only under forward bias conditions and the spin lifetime is estimated to be $\sim$ 45 nsec at 10 K. Since there are some reliable evidence,\cite{Croker,Lou2,Chan} the results by Lou {\it et al}.\cite{Lou} support that the Hanle-effect curves observed by two- and three-terminal methods are possible evidence for the electrical detection of spin-polarized electrons created in a GaAs conduction channel. Another was observed in Co/Al$_\text{2}$O$_\text{3}$/GaAs devices, reported by Tran {\it et al}.\cite{Tran} The observed Hanle-effect curves were significantly large and seen in both bias conditions. In this case, the estimated spin lifetime was $\sim$ 0.3 nsec at 10 K, which is significantly shorter than that reported by Lou {\it et al}.\cite{Lou} They discussed that such large spin signals and short spin lifetime originate from the two-step tunneling of spins through the localized state located in the vicinity of Al$_\text{2}$O$_\text{3}$/GaAs interface.\cite{Tran} Thus, by carefully observing Hanle-effect curves using the two- or three-terminal method, it seems that one can judge whether the spin accumulation is created in the conduction channel or in the oxide-related localized state.

For Si-based devices, Sasaki {\it et al.} recently clarified the three-terminal Hanle curves originating from the spin accumulation created not in the localized state but in a Si channel by comparison of the results measured by four-terminal nonlocal spin transport.\cite{Sasaki2} Sasaki {\it et al}. have used a metallic (heavily doped) Si channel and a contact with insulating tunnel barriers. Here, we focus on a device structure consisting of a semiconducting Si channel and a Schottky tunnel contact for semiconductor spintronic applications with a low parasitic resistance.\cite{Sugahara}

In this letter, we evidently show the Hanle-effect curves depending on the current-bias conditions in Si-based three-terminal lateral devices with a Schottky tunnel contact. In a low current-bias region, Hanle-effect curves can be observed only under forward bias conditions. We infer that the electrical detectability of the forward-biased contact is higher than that of the reverse-biased contact, as shown in Fe/GaAs devices reported previously.\cite{Lou} This is possible evidence for the detection of spin-polarized electrons created not in the localized state in the vicinity of the interface but in a Si channel.
\begin{figure}[t]
\includegraphics[width=6.5cm]{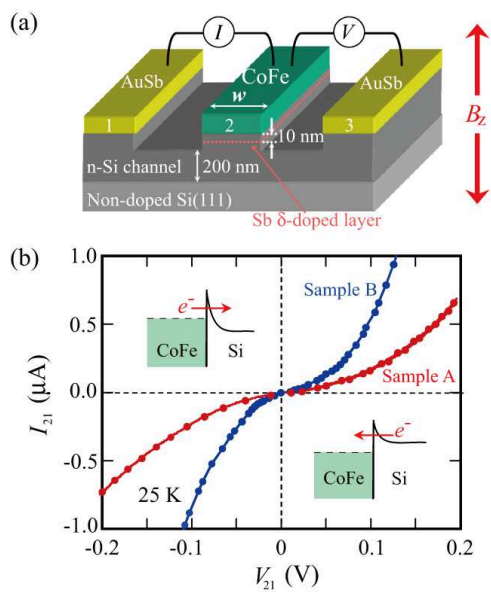}
\caption{(Color online) (a) Schematic diagram of a lateral three-terminal device with a CoFe/Si contact. (b) $I-V$ characteristics measured between contacts 1 and 2 for sample A ($w =$ 6 $\mu$m) and sample B ($w =$ 40 $\mu$m). }
\end{figure} 

Single-crystalline Co$_{60}$Fe$_{40}$ films with a thickness of 10 nm were grown on Si(111) by low-temperature molecular beam epitaxy (LT-MBE) at 60 $^{\circ}$C. Detailed formation procedures are described in Ref.\cite{Maeda}. Prior to the growth, we fabricated a phosphorus-doped n-Si(111) channel with a thickness of $\sim$ 200 nm on a non-doped  FZ-Si(111) substrate by using an ion implantation technique. The carrier densities estimated by Hall measurements were $n =$ $\sim$ 2$\times$10$^{18}$ cm$^{-3}$ at 300 K and $\sim$ 1$\times$10$^{15}$ cm$^{-3}$ at 25 K, respectively. Thus, our devices obviously have a semiconducting Si channel.\cite{Pearson,Morin} After the fabrication of the n-Si(111) channel, Sb $\delta$-doped n$^{+}$-Si layer (Sb $\sim$ 5$\times$10$^{19}$ cm$^{-3}$) with a thickness of 10 nm was grown by MBE.\cite{Ando} As a result, the realistic conduction band should have an energy profile with a Schottky-tunnel barrier, as shown in Ref. \cite{Song}. Here, since the doped Sb atoms can easily diffuse toward the surface because of the segregation during the growth, the carrier density near the interface can become $\sim$10$^{19}$ cm$^{-3}$ with some distributions. Conventional processes with photolithography, Ar$^{+}$ ion milling, and reactive ion etching were used to fabricate three-terminal lateral devices, called sample A ($w =$ 6 $\mu$m) and sample B ($w =$ 40 $\mu$m), with a CoFe/Si contact, illustrated in Fig. 1(a). The contact 2 of the sample A and sample B has a lateral dimension of 6 $\times$ 200 $\mu$m$^{2}$ or 40 $\times$ 200 $\mu$m$^{2}$, respectively. The AuSb ohmic contacts, labeled by 1 and 3, were formed and the Sb $\delta$-doped n$^{+}$-Si layer on the channel region was removed by the Ar$^{+}$ ion milling. The distance between the contacts 2 and 1 or 3 is $\sim$ 50 $\mu$m. The three-terminal Hanle measurements were performed by a dc method with the current-voltage configuration shown in Fig. 1(a) at 25 K. In the measurements, a small magnetic field perpendicular to the plane, $B_\text{Z}$, was applied after the magnetic moment of the contact 2 aligned parallel to the plane along the long axis of the contact. 
\begin{figure}[t]
\includegraphics[width=8.5cm]{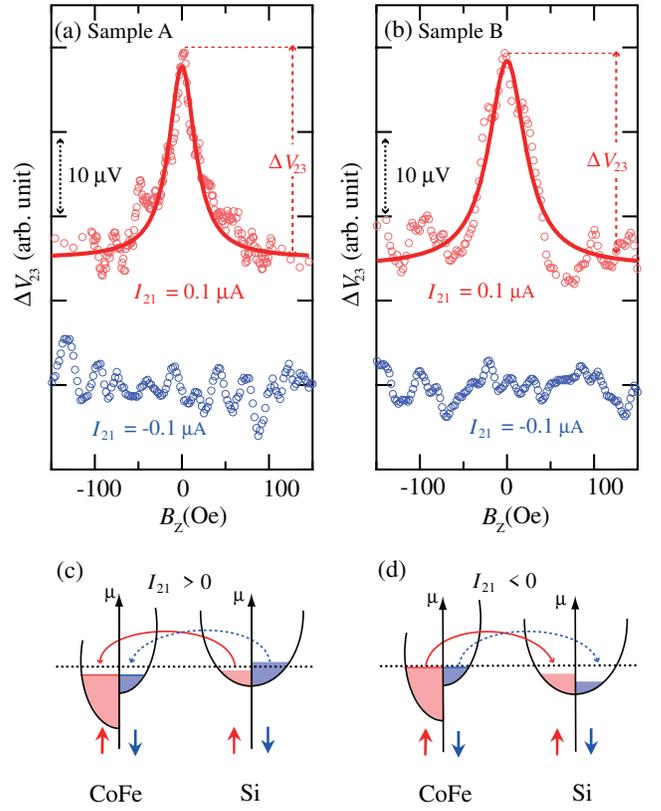}
\caption{(Color online) The voltage changes $\Delta$$V_\text{23}$ as a function of $B_\text{Z}$ at 25 K for (a) sample A and (b) sample B. The red and blue plots show $\Delta$$V_\text{23}$ for $I$$_\text{21} =$ 0.1 and - 0.1 $\mu$A. Schematic illustrations of spin accumulation in a Si conduction band by (c) spin extraction and (d) spin injection for $I$$_\text{21} > 0$ and $I$$_\text{21} < 0$, respectively. }
\end{figure}

Figure 1(b) shows two-terminal current-voltage ($I-V$) characteristics for sample A (red) and sample B (blue), measured between contacts 1 and 2 at 25 K. The electrons are extracted from and injected into, respectively, the Si conduction band for forward ($V_\text{21} > 0$) and reverse ($V_\text{21} < 0$) biases, as shown in the inset figures. For both samples nonlinear $I-V$ features can be seen. Though the contact width ($w$) of sample B is six times larger than that of sample A, the magnitude of the current obtained in sample B is smaller than that expected. Thus, the inhomogeneous doping profile is somewhat distributed in the film plane. Here, we have already confirmed that the $I-V$ characteristics are dominated by the Schottky tunnel barrier of the contact 2, i.e., the contact resistance (sample A $\sim$ 730 k$\Omega$, sample B $\sim$ 440 k$\Omega$) is much higher than the channel resistance ($\sim$ 30 k$\Omega$) at low temperatures. The zero-bias resistance of the contact 2 showed weak variations less than one order of magnitude with decreasing temperature from 300 to 25 K (not shown here). Thus, the transport mechanism is dominated by the tunneling conduction through the CoFe/Si interface.\cite{Hanbicki} 

The three-terminal voltage, $V$$_\text{23}$, as a function of $B_\text{Z}$, is shown in Figs. 2(a) and (b) for sample A and sample B, respectively. The measurements were performed at 25 K. The red and blue plots show $\Delta$$V_\text{23}$ for $I$$_\text{21} =$ 0.1 and - 0.1 $\mu$A, respectively, where a quadratic background voltage depending on $B_\text{Z}$ is subtracted from the raw data. For $I$$_\text{21} =$ 0.1 $\mu$A (red plot), a clear voltage drop ($\Delta$$V_\text{23}$) of 24 $\mu$V is observed when $B_\text{Z}$ increases from zero to $\pm$100 Oe. The voltage change is caused by the depolarization of the accumulated spins,\cite{Dash,Lou,Tran,Sasaki2} i.e., a Hanle-type spin precession is detected. On the other hand, for $I$$_\text{21} =$ -0.1 $\mu$A (blue plot), we can see no Hanle-like signal. The asymmetric bias dependence can tentatively be understood by a difference in the electrical detectability for the spin accumulation in a Si conduction band as illustrated in Figs. 2(c) and (d). Under forward bias [Fig. 2(c)], spin-polarized states for both Si and CoFe are located at the quasi Fermi level, so that the voltage change induced by the Hanle effect can be detected at the Schottky tunnel contact which dominates the $I-V$ characteristic in our devices. On the other hand, under reverse bias [Fig. 2(d)], the available state of Si near the quasi Fermi level remains unpolarized because of the fractional spin accumulation in a Si conduction band. As a result, the detectability for the voltage change induced by the Hanle effect is relatively low compared with the case under the forward bias. In a previous work of Fe/GaAs lateral devices reported by Lou {\it et al.},\cite{Lou} the voltage changes induced by the Hanle effect in a GaAs conduction channel were demonstrated only via a forward-biased Schottky tunnel contact. Our results observed in Fig. 2 are almost similar to the Fe/GaAs Schottky devices reported by Lou {\it et al.,}\cite{Lou} indicating that the detected Hanle-effect signals in Fig. 2 arise from the spin accumulation created in a Si conduction channel. Considering the theoretical equation of the Hanle-type spin precession described in Ref. 5, we can judge that the $\Delta$$V$$_\text{23}$ values obtained are predominantly resulting from the spin accumulation in the bulk Si channel with a carrier density of $\sim$ 10$^{15}$ cm$^{-3}$.\cite{comment} We could not detect voltage changes resulting from the spin accumulation in the heavily doped regions ($\sim$ 10$^{19}$ cm$^{-3}$) because the contribution of their voltage changes to $\Delta$$V$$_\text{23}$ is quite small in very low current-bias conditions. 
\begin{figure}[t]
\includegraphics[width=8cm]{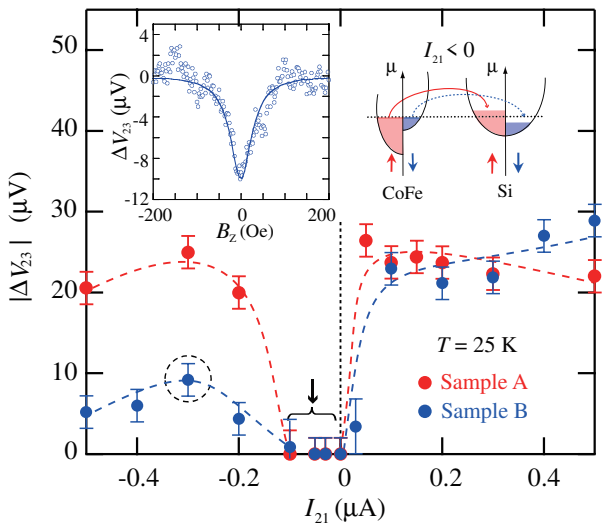}
\caption{(Color online) $|$$\Delta$$V_\text{23}$$|$ as a function of bias current $I$$_\text{21}$ for sample A (red) and sample B (blue) at 25 K. Left inset shows a $\Delta$$V_\text{23}-$$B_\text{Z}$ curve of sample B measured at $I$$_\text{21}= $-0.3 $\mu$A. Right inset illustrates the large spin accumulation in a Si conduction band in $I$$_\text{21} < 0$. }
\end{figure}

We can roughly estimate a lower limit of spin lifetime ($\tau_\text{S}$) using the Lorentzian function, $\Delta$$V_\text{23}$($B_\text{Z}$) $=$ $\Delta$$V_\text{23}$/[1+($\omega_\text{L}$$\tau_\text{S}$)$^{2}$], where $\omega_\text{L} =$ $g\mu_\text{B}$$B_\text{Z}$/$\hbar$ is the Lamor frequency, $g$ is the electron $g$-factor ($g =$ 2), $\mu_\text{B}$ is the Bohr magneton.\cite{Dash} The fitting curves (red solid curves) are shown in Figs. 2(a) and (b). As a consequence, the $\tau_\text{S}$ values of sample A and sample B are estimated to be $\sim$ 3.1 and $\sim$ 2.2 nsec, respectively. Recently, Sasaki {\it et al.} clarified that $\tau_\text{S}$ in a metallic Si channel ($n =$ 1$\times$10$^{19}$ cm$^{-3}$) at low temperatures was estimated to be 4 $\sim$ 9 nsec by the three-terminal methods.\cite{Sasaki2} Our obtained $\tau_\text{S}$ values are close to those reported by Sasaki {\it et al}.\cite{Sasaki2} This fact supports that our results originate from the spin accumulation not in the localized state in the vicinity of the CoFe/Si interface but in a Si conduction channel. Since our devices have a CoFe/Si interface and a very thin depletion layer induced by the $\delta$ doping, the spin accumulation created in the Si conduction channel can be detected through the Schottky tunnel contact. 

These $\tau_\text{S}$ values reported by Sasaki {\it et al}.\cite{Sasaki2} and in this study are significantly shorter than those ($\sim$ hundreds nsec) reported by Appelbaum {\it et al}.\cite{Appelbaum} Though the precise origin of the large difference in $\tau_\text{S}$ is still unclear, we speculate that the crystal quality and impurity density of the Si channels formed by an ion implantation technique are significantly different from those of the FZ-Si reported by Appelbaum {\it et al}.\cite{Appelbaum} For explaining the large difference in $\tau_\text{S}$, further investigations are required by using several devices with different impurity densities and so forth. 

Finally, we plots $|\Delta$$V_\text{23}|$ versus bias current ($I$$_\text{21}$) for both samples in Fig. 3. In low reverse $I$$_\text{21}$ (-0.1 $\mu$A $<$ $I$$_\text{21}$ $\le$ 0 $\mu$A), $|\Delta$$V_\text{23}|$ can not be detected (see arrow) for both samples, as discussed in the previous paragraph. However, we note that $|\Delta$$V_\text{23}|$ can be evidently seen in $I$$_\text{21} \le$ -0.1 $\mu$A. For example, we show a $\Delta$$V_\text{23}-$$B_\text{Z}$ curve for sample B at $I$$_\text{21} =$ -0.3 $\mu$A in the left inset. In this study the sign reversal of $\Delta$$V_\text{23}$ occurs only when the current polarity changes. Thus, this feature can be explained as a consequence of the enhancement in the spin accumulation in the Si channel by increasing injected spins. That is, with increasing spin accumulation, the available state of the Si conduction band near the quasi Fermi level can become spin-polarized, as shown in the right inset. Therefore, the electrical detectability of the three-terminal method for spin accumulation can vary with changing bias current. These results may also support the electrical detection of spin-polarized electrons created in a semiconducting Si channel. 

In summary, we have demonstrated evident bias-current dependence of the spin accumulation in a semiconducting Si channel at 25 K by measuring the Hanle-effect signals. We have observed that the electrical detectability of the Hanle-effect signals at the forward-biased contact is higher than that at the reverse-biased contact. This work indicates possible evidence for the detection of spin-polarized electrons created not in the localized state in the vicinity of the interface but in a Si channel.

This work was partly supported by PRESTO-JST and STARC. One of the authors (Y.A.) acknowledges JSPS Research Fellowships for Young Scientists.



\begin{thebibliography}{11}
\bibitem{Lou}
X. Lou, C. Adelmann, M. Furis, S. A. Crooker, C. J. Palmstr\o m, and P. A. Crowell, Phys. Rev. Lett. {\bf 96}, 176603 (2006).
\bibitem{Appelbaum}
I. Appelbaum, B. Huang, and D. J. Monsma, Nature {\bf 447}, 295 (2007); H. Jang and I. Appelbaum, Phys. Rev. Lett. {\bf 103}, 117202 (2009).
\bibitem{Tran}
M. Tran, H. Jaffr\`es, C. Deranlot, J. -M. George, A. Fert, A. Miard, and A. Lema\^{\i}tre, Phys. Rev. Lett. {\bf 102}, 036601 (2009).
\bibitem{Dash}
S. P. Dash, S. Sharma, R. S. Patel1, M. P. Jong, and R. Jansen, Nature (London) {\bf 462}, 491 (2009); S. P. Dash, S. Sharma, J. C. Le Breton, and R. Jansen, Proc. SPIE 7760, 77600J (2010).
\bibitem{Sasaki2}
T. Sasaki, T. Oikawa, M. Shiraishi, Y. Suzuki, and K. Noguchi, Appl. Phys. Lett. {\bf 98}, 012508 (2011).
\bibitem{Croker}
S. A. Crooker, M. Furis, X. Lou, C. Adelmann, D. L. Smith, C. J. Palmstr\o m, and P. A. Crowell, Science {\bf 309}, 2191 (2005).
\bibitem{Lou2}
X. Lou, C. Adelmann, S. A. Crooker, E. S. Garlid, J. Zhang, S. M. Reddy, S. D. Flexner, C. J. Palmstr\o m, and P. A. Crowell, Nat. Phys. {\bf 3}, 197 (2007).
\bibitem{Chan}
M. K. Chan, Q. O. Hu, J. Zhang, T. Kondo, C. J. Palmstr\o m, and P. A. Crowell, Phys. Rev. B {\bf 80}, 161206(R) (2009).
\bibitem{Sugahara}
S. Sugahara and M. Tanaka, Appl. Phys. Lett. {\bf 84}, 2307 (2004).
\bibitem{Maeda}
Y. Maeda, K. Hamaya, S. Yamada, Y. Ando, K. Yamane, and M. Miyao, Appl. Phys. Lett. {\bf 97}, 192501 (2010).
\bibitem{Pearson}
G. L. Pearson and J. Bardeen, Phys. Rev. {\bf 75}, 865 (1949).
\bibitem{Morin}
F. J. Morin and J. P. Maita, Phys. Rev. {\bf 96}, 28 (1954).
\bibitem{Ando}
Y. Ando, K. Hamaya, K. Kasahara, Y. Kishi, K. Ueda, K. Sawano, T. Sadoh, and M. Miyao, Appl. Phys. Lett. {\bf 94},182105 (2009); K. Hamaya, Y. Ando, T. Sadoh, and M. Miyao, Jpn. J. Appl. Phys. {\bf 50}, 010101 (2011).
\bibitem{Song}
Y. Song and H. Dery, Phys. Rev. B {\bf 81}, 045321 (2010).
\bibitem{Hanbicki}
A. T. Hanbicki, O. M. J. vanft Erve, R. Magno, G. Kioseoglou, C. H. Li, B. T. Jonker, G. Itskos, R. Mallory, M. Yasar, and A. Petrou, Appl. Phys. Lett. {\bf 82}, 4092 (2003).
\bibitem{comment}
If we use the values of the spin diffusion length and conductivity of the heavily doped Si,\cite{Sasaki2} the estimated spin polarization can easily become more than 1.0, inconsistent with the bulk spin polarization of CoFe electrodes. 

\end{thebibliography}
\end{document}